\documentclass[conference]{IEEEtran}

\usepackage{cite}
\usepackage[T1]{fontenc} 
\ifCLASSINFOpdf
\else
\fi
\usepackage{bm,color,soul}
\usepackage{amsmath, amsthm, amssymb}
\theoremstyle{plain}

\usepackage[pdftex]{graphics}
\theoremstyle{definition}

\usepackage{xspace}
\usepackage[nolist,printonlyused]{acronym}    
\usepackage{amssymb}
\usepackage{graphicx}
\usepackage{amsmath}
\usepackage{mathtools}
\usepackage{graphicx,color} 
\usepackage{amsmath, amsfonts, amssymb, amsbsy,nccmath} 
\usepackage{algorithm} 
\usepackage{enumerate} 
\usepackage{algorithmic} 
\usepackage{lipsum} 
\usepackage{mathtools}
\usepackage{dsfont} 
\usepackage[inline]{enumitem}
\usepackage{amsmath}
\usepackage[cmintegrals]{newtxmath}
\usepackage{color,soul}
\usepackage{stfloats}  
\usepackage{tabularx}
\usepackage{lipsum}
\usepackage{mwe}
\usepackage{amsmath}
\usepackage{longtable}
\usepackage{subcaption}

\newcommand{\bmA}{\mathbf A}
\newcommand{\bma}{\mathbf a}
\newcommand{\bmI}{\mathbf I}

\newcommand{\bmB}{\mathbf B}
\newcommand{\bmC}{\mathbf C}
\newcommand{\bmD}{\mathbf D}
\newcommand{\bmE}{\mathbf E}

\newcommand{\bmH}{\mathbf H}
\newcommand{\bmF}{\mathbf F}

\newcommand{\bms}{\mathbf s}
\newcommand{\bmd}{\mathbf d}
\newcommand{\bmV}{\mathbf V}

\newcommand{\bmy}{\mathbf y}

\newcommand{\bmR}{\mathbf R}

\newcommand{\bmW}{\mathbf W}

\newcommand{\bmq}{\mathbf q}

\newcommand{\bmn}{\mathbf n}

\usepackage[T1]{fontenc}
\usepackage[utf8]{inputenc}
\usepackage{authblk}

\begin{document}
\title{Full-Duplex-Enabled Joint Communications and Sensing with Reconfigurable Intelligent Surfaces}
\author{Chandan~Kumar~Sheemar\IEEEauthorrefmark{1}, George C. Alexandropoulos\IEEEauthorrefmark{2}, Dirk Slock\IEEEauthorrefmark{3}, 
Jorge Querol\IEEEauthorrefmark{1}, 
and Symeon Chatzinotas\IEEEauthorrefmark{1}\\
\IEEEauthorrefmark{1}SnT, University of Luxembourg, emails: \{chandankumar.sheemar, jorge.querol, symeon.chatzinotas\}@uni.lu,\\
\IEEEauthorrefmark{2} National and Kapodistrian University of Athens, Greece, email\{alexandg\}@di.uoa.gr\\
\IEEEauthorrefmark{3}EURECOM, Sophia Antipolis, France, email:\{slock\}@eurecom.fr
}

\maketitle 
\begin{acronym}
  \acro{2G}{Second Generation}
  \acro{3G}{3$^\text{rd}$~Generation}
  \acro{3GPP}{3$^\text{rd}$~Generation Partnership Project}
  \acro{4G}{4$^\text{th}$~Generation}
  \acro{5G}{5$^\text{th}$~Generation}
  \acro{AA}{Antenna Array}
  \acro{AC}{Admission Control}
  \acro{ACL}{adjacent channel leakage}
  \acro{AD}{Attack-Decay}
  \acro{ADC}{analog-to-digital converter}
  \acro{ADSL}{Asymmetric Digital Subscriber Line}
  \acro{AHW}{Alternate Hop-and-Wait}
  \acro{AMC}{Adaptive Modulation and Coding}
	\acro{AP}{Access Point}
  \acro{APA}{Adaptive Power Allocation}
  \acro{AR}{autoregressive}
  \acro{ARMA}{Autoregressive Moving Average}
  \acro{ATES}{Adaptive Throughput-based Efficiency-Satisfaction Trade-Off}
  \acro{AWGN}{additive white Gaussian noise}
  \acro{A/D}{analog and digital}
  \acro{BB}{branch and bound}
  \acro{BD}{block diagonalization}
  \acro{BER}{bit error rate}
  \acro{BF}{Best Fit}
  \acro{BLER}{BLock Error Rate}
  \acro{BPC}{Binary power control}
  \acro{BPSK}{binary phase-shift keying}
  \acro{BPA}{Best \ac{PDPR} Algorithm}
  \acro{BRA}{Balanced Random Allocation}
  \acro{BS}{base station}
  \acro{CAP}{Combinatorial Allocation Problem}
  \acro{CAPEX}{Capital Expenditure}
  \acro{CBF}{Coordinated Beamforming}
  \acro{CBR}{Constant Bit Rate}
  \acro{CBS}{Class Based Scheduling}
  \acro{CC}{Congestion Control}
  \acro{CDF}{Cumulative Distribution Function}
  \acro{CDMA}{Code-Division Multiple Access}
  \acro{CL}{Closed Loop}
  \acro{CLI}{cross-link interference}
  \acro{CLPC}{Closed Loop Power Control}
  \acro{CNR}{Channel-to-Noise Ratio}
  \acro{CPA}{Cellular Protection Algorithm}
  \acro{CPICH}{Common Pilot Channel}
  \acro{CoMP}{Coordinated Multi-Point}
  \acro{CQI}{Channel Quality Indicator}
  \acro{CRM}{Constrained Rate Maximization}
	\acro{CRN}{Cognitive Radio Network}
  \acro{CS}{Coordinated Scheduling}
  \acro{CSI}{channel state information}
  \acro{CSIR}{channel state information at the receiver}
  \acro{CSIT}{channel state information at the transmitter}
  \acro{CUE}{cellular user equipment}
  \acro{DAC}{digital-to-analog converter}
  \acro{D2D}{device-to-device}
  \acro{DCA}{Dynamic Channel Allocation}
  \acro{DE}{Differential Evolution}
  \acro{DFT}{Discrete Fourier Transform}
  \acro{DIST}{Distance}
  \acro{DL}{downlink}
  \acro{DMA}{Double Moving Average}
	\acro{DMRS}{Demodulation Reference Signal}
  \acro{D2DM}{D2D Mode}
  \acro{DMS}{D2D Mode Selection}
  \acro{DOCSIS}{Data Over Cable Service Interface Specification}
  \acro{DPC}{Dirty Paper Coding}
  \acro{DRA}{Dynamic Resource Assignment}
  \acro{DSA}{Dynamic Spectrum Access}
  \acro{DSM}{Delay-based Satisfaction Maximization}
  \acro{EBD}{electrical balance duplexer}
  \acro{ECC}{Electronic Communications Committee}
  \acro{EFLC}{Error Feedback Based Load Control}
  \acro{EI}{Efficiency Indicator}
  \acro{eNB}{Evolved Node B}
  \acro{EPA}{Equal Power Allocation}
  \acro{EPC}{Evolved Packet Core}
  \acro{EPS}{Evolved Packet System}
  \acro{E-UTRAN}{Evolved Universal Terrestrial Radio Access Network}
  \acro{ES}{Exhaustive Search}
  \acro{FD}{full-duplex}
  \acro{FDD}{frequency division duplexing}
  \acro{FDM}{Frequency Division Multiplexing}
  \acro{FER}{Frame Erasure Rate}
  \acro{FF}{Fast Fading}
  \acro{FSB}{Fixed Switched Beamforming}
  \acro{FST}{Fixed SNR Target}
  \acro{FTP}{File Transfer Protocol}
  \acro{GA}{Genetic Algorithm}
  \acro{GBR}{Guaranteed Bit Rate}
  \acro{GLR}{Gain to Leakage Ratio}
  \acro{GOS}{Generated Orthogonal Sequence}
  \acro{GPL}{GNU General Public License}
  \acro{GRP}{Grouping}
  \acro{HARQ}{Hybrid Automatic Repeat Request}
  \acro{HD}{half-duplex}
  \acro{HMS}{Harmonic Mode Selection}
  \acro{HOL}{Head Of Line}
  \acro{HSDPA}{High-Speed Downlink Packet Access}
  \acro{HSPA}{High Speed Packet Access}
  \acro{HTTP}{HyperText Transfer Protocol}
  \acro{ICMP}{Internet Control Message Protocol}
  \acro{ICI}{Intercell Interference}
  \acro{ID}{Identification}
  \acro{IETF}{Internet Engineering Task Force}
  \acro{ILP}{Integer Linear Program}
  \acro{JRAPAP}{Joint RB Assignment and Power Allocation Problem}
  \acro{UID}{Unique Identification}
  \acro{IBFD}{in-band full-duplex}
  \acro{IID}{Independent and Identically Distributed}
  \acro{IIR}{Infinite Impulse Response}
  \acro{ILP}{Integer Linear Problem}
  \acro{IMT}{International Mobile Telecommunications}
  \acro{INV}{Inverted Norm-based Grouping}
  \acro{IoT}{Internet of Things}
  \acro{IP}{Internet Protocol}
  \acro{IPv6}{Internet Protocol Version 6}
  \acro{I/Q}{in-phase/quadrature}
  \acro{IRS}{Intelligent Reflective Surface}
  \acro{ISAC}{integrated sensing and communications}
  \acro{ISD}{Inter-Site Distance}
  \acro{ISI}{Inter Symbol Interference}
  \acro{ITU}{International Telecommunication Union}
  \acro{JCAS}{joint communications and sensing}
  \acro{JOAS}{Joint Opportunistic Assignment and Scheduling}
  \acro{JOS}{Joint Opportunistic Scheduling}
  \acro{JP}{Joint Processing}
	\acro{JS}{Jump-Stay}
    \acro{KF}{Kalman filter}
  \acro{KKT}{Karush-Kuhn-Tucker}
  \acro{L3}{Layer-3}
  \acro{LAC}{Link Admission Control}
  \acro{LA}{Link Adaptation}
  \acro{LC}{Load Control}
  \acro{LNA}{low-noise amplifier}
  \acro{LOS}{Line of Sight}
  \acro{LP}{Linear Programming}
  \acro{LS}{least squares}
  \acro{LTE}{Long Term Evolution}
  \acro{LTE-A}{LTE-Advanced}
  \acro{LTE-Advanced}{Long Term Evolution Advanced}
  \acro{M2M}{Machine-to-Machine}
  \acro{MAC}{Medium Access Control}
  \acro{MANET}{Mobile Ad hoc Network}
  \acro{MC}{Modular Clock}
  \acro{MCS}{Modulation and Coding Scheme}
  \acro{MDB}{Measured Delay Based}
  \acro{MDI}{Minimum D2D Interference}
  \acro{MF}{Matched Filter}
  \acro{MG}{Maximum Gain}
  \acro{MH}{Multi-Hop}
  \acro{MIMO}{multiple-input multiple-output}
  \acro{MINLP}{Mixed Integer Nonlinear Programming}
  \acro{MIP}{Mixed Integer Programming}
  \acro{MISO}{Multiple Input Single Output}
  \acro{ML}{maximum likelihood}
  \acro{MLWDF}{Modified Largest Weighted Delay First}
  \acro{mmWave}{millimeter wave}
  \acro{MME}{Mobility Management Entity}
  \acro{MMSE}{minimum mean squared error}
  \acro{MOS}{Mean Opinion Score}
  \acro{MPF}{Multicarrier Proportional Fair}
  \acro{MRA}{Maximum Rate Allocation}
  \acro{MR}{Maximum Rate}
  \acro{MRC}{maximum ratio combining}
  \acro{MRT}{Maximum Ratio Transmission}
  \acro{MRUS}{Maximum Rate with User Satisfaction}
  \acro{MS}{mobile station}
  \acro{MSE}{mean squared error}
  \acro{MSI}{Multi-Stream Interference}
  \acro{MTC}{Machine-Type Communication}
  \acro{MTSI}{Multimedia Telephony Services over IMS}
  \acro{MTSM}{Modified Throughput-based Satisfaction Maximization}
  \acro{MU-MIMO}{multiuser multiple input multiple output}
  \acro{MU}{multi-user}
  \acro{NAS}{Non-Access Stratum}
  \acro{NB}{Node B}
  \acro{NE}{Nash equilibrium}
  \acro{NCL}{Neighbor Cell List}
  \acro{NLP}{Nonlinear Programming}
  \acro{NLOS}{Non-Line of Sight}
  \acro{NMSE}{Normalized Mean Square Error}
  \acro{NORM}{Normalized Projection-based Grouping}
  \acro{NP}{Non-Polynomial Time}
  \acro{NR}{New Radio}
  \acro{NRT}{Non-Real Time}
  \acro{NSPS}{National Security and Public Safety Services}
  \acro{O2I}{Outdoor to Indoor}
  \acro{OFDMA}{orthogonal frequency division multiple access}
  \acro{OFDM}{orthogonal frequency division multiplexing}
  \acro{OFPC}{Open Loop with Fractional Path Loss Compensation}
  \acro{O2I}{Outdoor-to-Indoor}
  \acro{OL}{Open Loop}
  \acro{OLPC}{Open-Loop Power Control}
  \acro{OL-PC}{Open-Loop Power Control}
  \acro{OPEX}{Operational Expenditure}
  \acro{ORB}{Orthogonal Random Beamforming}
  \acro{JO-PF}{Joint Opportunistic Proportional Fair}
  \acro{OSI}{Open Systems Interconnection}
  \acro{PAIR}{D2D Pair Gain-based Grouping}
  \acro{PAPR}{Peak-to-Average Power Ratio}
  \acro{PBCH}{physical broadcast channel}
  \acro{P2P}{Peer-to-Peer}
  \acro{PC}{Power Control}
  \acro{PCI}{Physical Cell ID}
  \acro{PDF}{Probability Density Function}
  \acro{PDPR}{pilot-to-data power ratio}
  \acro{PER}{Packet Error Rate}
  \acro{PF}{Proportional Fair}
  \acro{P-GW}{Packet Data Network Gateway}
  \acro{PL}{Pathloss}
  \acro{PMN}{Perceptive Mobile Network}
  \acro{PPR}{pilot power ratio}
  \acro{PRB}{physical resource block}
  \acro{PROJ}{Projection-based Grouping}
  \acro{ProSe}{Proximity Services}
  \acro{PS}{Packet Scheduling}
  \acro{PSAM}{pilot symbol assisted modulation}
  \acro{PSO}{Particle Swarm Optimization}
  \acro{PZF}{Projected Zero-Forcing}
  \acro{QAM}{Quadrature Amplitude Modulation}
  \acro{QoS}{Quality of Service}
  \acro{QPSK}{Quadri-Phase Shift Keying}
  \acro{RAISES}{Reallocation-based Assignment for Improved Spectral Efficiency and Satisfaction}
  \acro{RAN}{Radio Access Network}
  \acro{RA}{Resource Allocation}
  \acro{RAT}{Radio Access Technology}
  \acro{RATE}{Rate-based}
  \acro{RB}{resource block}
  \acro{RBG}{Resource Block Group}
  \acro{REF}{Reference Grouping}
  \acro{RF}{radio frequency}
  \acro{RIS}{reconfigurable intelligent surface}
  \acro{RLC}{Radio Link Control}
  \acro{RM}{Rate Maximization}
  \acro{RNC}{Radio Network Controller}
  \acro{RND}{Random Grouping}
  \acro{RRA}{Radio Resource Allocation}
  \acro{RRM}{Radio Resource Management}
  \acro{RSCP}{Received Signal Code Power}
  \acro{RSRP}{Reference Signal Receive Power}
  \acro{RSRQ}{Reference Signal Receive Quality}
  \acro{RR}{Round Robin}
  \acro{RRC}{Radio Resource Control}
  \acro{RSSI}{Received Signal Strength Indicator}
  \acro{RT}{Real Time}
  \acro{RU}{Resource Unit}
  \acro{RUNE}{RUdimentary Network Emulator}
  \acro{RV}{Random Variable}
  \acro{RX}{receiver}
  \acro{SAC}{Session Admission Control}
  \acro{SBFD}{sub-band full-duplex}
  \acro{SCM}{Spatial Channel Model}
  \acro{SC-FDMA}{Single Carrier - Frequency Division Multiple Access}
  \acro{SD}{Soft Dropping}
  \acro{S-D}{Source-Destination}
  \acro{SDPC}{Soft Dropping Power Control}
  \acro{SDMA}{Space-Division Multiple Access}
  \acro{SER}{Symbol Error Rate}
  \acro{SES}{Simple Exponential Smoothing}
  \acro{S-GW}{Serving Gateway}
  \acro{SI}{self-interference}
  \acro{SINR}{signal-to-interference-plus-noise ratio}
  \acro{SIC}{self-interference cancellation}
  \acro{SIP}{Session Initiation Protocol}
  \acro{SISO}{single-input single-output}
  \acro{SIMO}{Single Input Multiple Output}
  \acro{SIR}{signal-to-interference ratio}
  \acro{SLNR}{Signal-to-Leakage-plus-Noise Ratio}
  \acro{SMA}{Simple Moving Average}
  \acro{SNR}{signal-to-noise ratio}
  \acro{SORA}{Satisfaction Oriented Resource Allocation}
  \acro{SORA-NRT}{Satisfaction-Oriented Resource Allocation for Non-Real Time Services}
  \acro{SORA-RT}{Satisfaction-Oriented Resource Allocation for Real Time Services}
  \acro{SPF}{Single-Carrier Proportional Fair}
  \acro{SRA}{Sequential Removal Algorithm}
  \acro{SRS}{Sounding Reference Signal}
  \acro{STAR}{simultaneous transmit-and-receive}
  \acro{SU-MIMO}{single-user multiple input multiple output}
  \acro{SU}{Single-User}
  \acro{SVD}{Singular Value Decomposition}
  \acro{TCP}{Transmission Control Protocol}
  \acro{TDD}{time division duplexing}
  \acro{TDMA}{Time Division Multiple Access}
  \acro{TETRA}{Terrestrial Trunked Radio}
  \acro{TP}{Transmit Power}
  \acro{TPC}{Transmit Power Control}
  \acro{TTI}{Transmission Time Interval}
  \acro{TTR}{Time-To-Rendezvous}
  \acro{TSM}{Throughput-based Satisfaction Maximization}
  \acro{TU}{Typical Urban}
  \acro{TX}{transmit}
  \acro{RX}{receive}
  \acro{UE}{User Equipment}
  \acro{UEPS}{Urgency and Efficiency-based Packet Scheduling}
  \acro{UL}{uplink}
  \acro{UMTS}{Universal Mobile Telecommunications System}
  \acro{URI}{Uniform Resource Identifier}
  \acro{URM}{Unconstrained Rate Maximization}
  \acro{UT}{user terminal}
  \acro{VR}{Virtual Resource}
  \acro{VoIP}{Voice over IP}
  \acro{WAN}{Wireless Access Network}
  \acro{WCDMA}{Wideband Code Division Multiple Access}
  \acro{WF}{Water-filling}
  \acro{WiMAX}{Worldwide Interoperability for Microwave Access}
  \acro{WINNER}{Wireless World Initiative New Radio}
  \acro{WLAN}{Wireless Local Area Network}
  \acro{WMPF}{Weighted Multicarrier Proportional Fair}
  \acro{WPF}{Weighted Proportional Fair}
  \acro{WSN}{Wireless Sensor Network}
  \acro{WWW}{World Wide Web}
  \acro{XIXO}{(Single or Multiple) Input (Single or Multiple) Output}
  \acro{XDD}{cross-division duplex}
  \acro{ZF}{zero-forcing}
  \acro{ZMCSCG}{Zero Mean Circularly Symmetric Complex Gaussian}
\end{acronym}

\begin{abstract}  
The full-duplex (FD) technology has the potential to radically evolve wireless systems, facilitating the integration of both communications and radar functionalities into a single device, thus, enabling joint communication and sensing (JCAS). In this paper, we present a novel approach for JCAS that incorporates a reconfigurable intelligent surface (RIS) in the near-field of an FD multiple-input multiple-output (MIMO) node, which is jointly optimized with the digital beamformers to enable JSAC and efficiently handle self-interference (SI). We propose a novel problem formulation for FD MIMO JCAS systems to jointly minimize the total received power at the FD node's radar receiver, while maximizing the sum rate of downlink communications subject to a Cram\'{e}r-Rao bound (CRB) constraint. In contrast to the typically used CRB in the relevant literature, we derive a novel, more accurate, target estimation bound that fully takes into account the RIS deployment. The considered problem is solved using alternating optimization, which is guaranteed to converge to a local optimum. The simulation results demonstrate that the proposed scheme achieves significant performance improvement both for communications and sensing. It is showcased that, jointly designing the FD MIMO beamformers and the RIS phase configuration to be SI aware can significantly loosen the requirement for additional SI cancellation.
\end{abstract}


\IEEEpeerreviewmaketitle

\section{Introduction}
Full-duplex (FD) systems refer to wireless systems in which the same frequency band is used for both transmitting and receiving signals simultaneously, allowing for bi-directional communications\cite{sheemar2022practical}. This is in contrast to traditional half-duplex systems, in which a separate frequency band is used for transmitting and receiving signals, and the system can only perform one function at a time \cite{liu2015band}. FD systems have the potential to significantly improve the performance and capacity of wireless networks, by allowing for more efficient use of the available spectrum, thus, enabling higher data rates. 

Self-interference (SI) is a major challenge for FD systems and SI cancellation (SIC) is viable to make FD a reality \cite{alexandropoulos2020full,sheemar2021hybrid}. Due to simultaneous transmission and reception, FD is currently receiving significant interest as it has the potential to enable joint communication and sensing (JCAS), implying that next-generation base stations (BSs) and radar functionalities could be integrated into a single device. In \cite{ISAC2022}, a novel JSAC system transceiver design for an FD MIMO \ac{BS} equipped with hybrid analog and digital beamformers is presented.   Recently, reconfigurable intelligent surfaces (RISs) JCAS has gained significant interest due to its potential to increase communications and sensing performance \cite{barneto2022millimeter}. In \cite{wang2021joint}, RIS-assisted JCAS under the Cram\'{e}r-Rao bound (CRB) constraint was derived. However, the derived CRB neglected the RIS contribution in enhancing the sensing performance. In \cite{9852716}, RIS-assisted JCAS to maximize the sum rate under a radar beam-pattern similarity constraint was investigated. In \cite{song2022intelligent}, a CRB minimization-based beamforming design for non-line-of-sight (LoS) RIS-assisted JCAS was studied. Note that the designs presented above do not consider the effect of SI on JCAS which may overwhelm the receiver. Moreover, the CRB derived in \cite{wang2021joint} did not consider any effect of the RIS and, in \cite{song2022intelligent}, only the CRB for the non-LoS sensing case was derived.

In this paper, we focus on the optimization of an FD JCAS system comprising one MIMO downlink (DL) user and $1$ target to be detected by the FD node's radar receiver. Firstly, we derive the exact CRB for RIS-assisted FD JCAS, by considering both the LoS and the non-LoS contributions. Then, we propose a novel formulation to jointly minimizing the effective SI power received in uplink (UL) at the radar receiver while maximizing the sum rate for the DL user, subject to the derived exact CRB constraint. However, we remark that the optimization of RIS with exact CRB constraint is extremely challenging and would require significant additional space to be elaborated in detail. Due to space limitations, we consider imposing the CRB constraint only on the digital beamformer. The considered joint optimization problem is transformed into its equivalent minimization of the mean squared error (MMSE) problem \cite{christensen2008weighted}, which results to be composed of two terms: the SI power at the radar and the MSE at the DL user. A novel alternating optimization method is devised to minimize the overall objective function. Our goal is to tackle SI with both the digital beamformer at the MIMO transmitter and the passive beamforming offered by the RIS, hence, enabling close to ideal FD operation for JCAS.

Simulation results show that the proposed scheme achieves significant performance gain compared to the conventional JCAS scheme with no RIS. Moreover, we also show that RIS can be beneficial to improve both the communications performance and the sensing performance, while also assisting in lowering the SI.

The rest of the paper is organized as follows. Section \ref{sistema} presents the system model and the exact CRB for FD JCAS. Section \ref{algo} presents a novel joint beamforming design. Finally, Sections \ref{risultati} and \ref{conclusioni} present simulations results and conclusions, respectively.\footnote{\emph{Notations:}   Boldface lower and upper case characters denote vectors and matrices, respectively. $\mathbb{E}\{\cdot\}$, $\mbox{Tr}\{\cdot\}$, and $\bmI$ denote expectation, trace, and identity matrix, respectively.  The superscripts $(\cdot)^T$ and $(\cdot)^H$ denote transpose
and conjugate-transpose (Hermitian) operators, respectively.}

 \section{System Model} \label{sistema}

\begin{figure}
     \centering
\includegraphics[width=8cm,height=5cm]{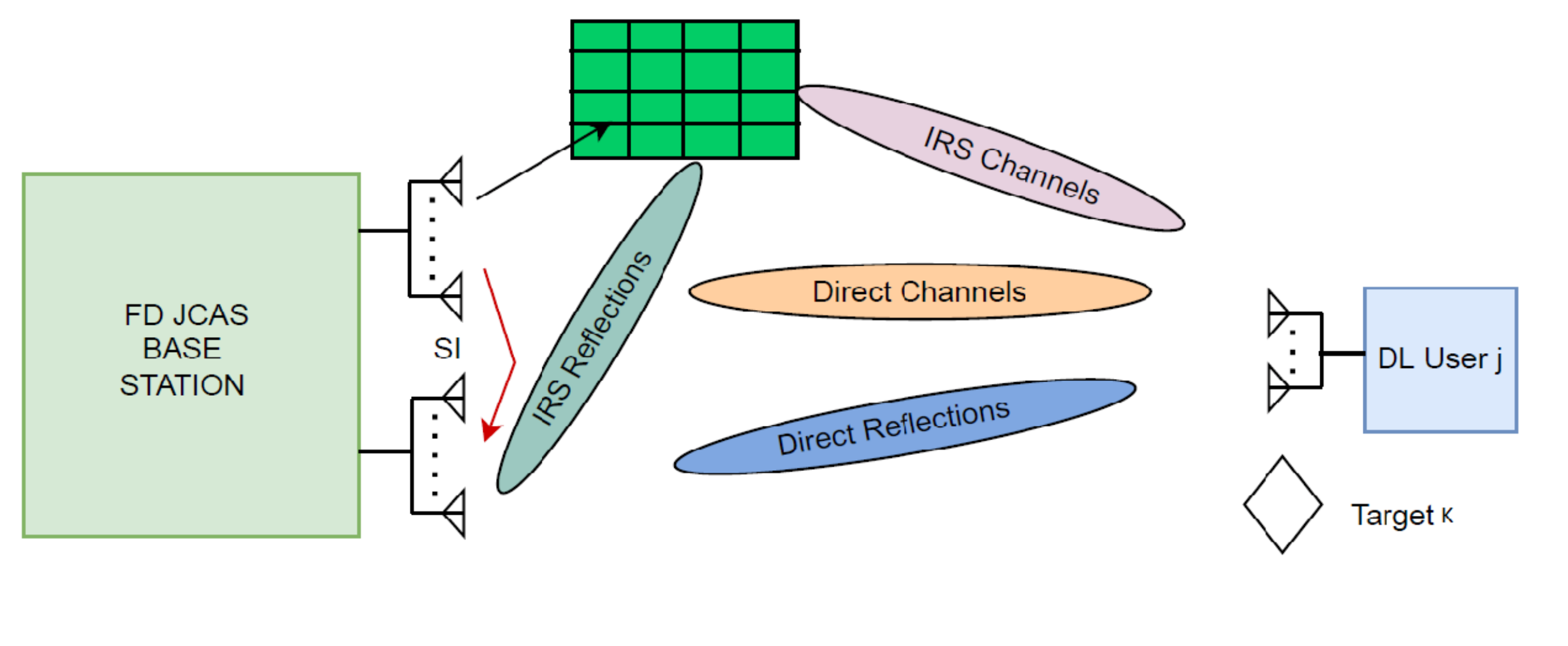}
     \caption{The proposed FD MIMO JCAS system with RIS.}
\label{fig:my_label}\vspace{-4mm}
 \end{figure}

We consider an FD JCAS system consisting of one FD MIMO BS $b$ equipped with $M_b$ and $N_b$ transmit and receive antennas, respectively, which communicates with one MIMO DL user $j$ having $N_j$ antenna elements, as shown in Fig. \ref{fig:my_label}. The DL signal is also used to detect targets/scatterers (via their induced reflections) randomly distributed within the communication/sensing environment at the radar receiver of the FD MIMO node. We  assume that the BS is assisted by a RIS placed within its near-field region, whose role is to jointly improve communication and sensing performance while offering strong SIC. Let $\bmV_{j} \in \mathbb{C}^{N_j \times M_j}$ denote the digital beamformer for the unit-variance data stream $\bms_j \in \mathbb{C}^{d_j \times 1}$ intended for the DL user $j$. The RIS is assumed to have $R$ rows and $C$ columns of reflection-tunable elements, whose phase response is denoted by the diagonal matrix $\mathbf{\Phi}_i \in \mathbb{C}^{RC \times RC}$, containing $\mathbf{\phi}_i \in \mathbb{C}^{RC \times 1}$ on its main diagonal. 
 
Let $\bmH_{j,b} \in \mathbb{C}^{N_j \times M_b}$ and $ \bmH_{j,i} \in \mathbb{C}^{N_j\times R C}$ denote the channels from the BS and the RIS to the DL user $j$, respectively. The channel from the RIS to the BS and from the BS to RIS are denoted with $\bmH_{b,i} \in \mathbb{C}^{N_b \times R C}$ and $\bmH_{i,b} \in \mathbb{C}^{R C \times M_b }$, respectively. Let $\bmH_{b,b} \in \mathbb{C}^{N_b \times M_b}$ denote the SI channel for the MIMO FD JCAS node $b$ which can be decomposed as  $\mathbf{H}_{b,b} =  \mathbf{H}_{b,b}^{l} + \mathbf{H}_{b,b}^{r}$,
where $\mathbf{H}_{b,b}^{l} $ and $\mathbf{H}_{b,b}^{r}$ denote 
the LoS and non-LoS contributions for the SI channel, respectively.  
Since the transmit and receive antennas of the FD JCAS are in the near-field, we consider a spherical wavefront and model each element of $\mathbf{H}_{b,b}^{l}$ as
\begin{equation} \label{SI_LOS_model} \vspace{-2mm}
     \mathbf{H}_{b,b}^{l}(m,n) = \frac{\rho}{r_{m,n}} e^{-i 2 \pi \frac{d_{m,n}}{\lambda}}, \quad \forall m, n, \vspace{-1mm}
\end{equation}
where the scalars $\lambda$ and $\rho$ denote the wavelength and the power normalization constant to assure $\mathbb{E}(||\mathbf{H}_{b,b}^{l}||_F^2)=M_b N_b$, respectively, and the scalar $d_{m,n}$ denotes the distance between $m$-th receive and $n$-th transmit antenna. Depending on the size of RIS, the channels $\bmH_{b,i}$ and $\bmH_{i,b}$ can also be in the near-field. Therefore, we consider modelling them similarly as \eqref{SI_LOS_model}. We assume perfect CSI, which can be obtained by exploiting channel reciprocity via time division duplexing (TDD).

\subsection{Communication Model}
Let $\bmy_j$ denote the received signal at the DL user $j$, which can be written as
\vspace{-2mm}
\begin{equation}
    \bmy_j =  ( \bmH_{j,b} + \bmH_{j,i} \mathbf{\Phi}\bmH_{i,b}) \bmV_j \bms_j  + \bmn_j,
\end{equation}
 where $\bmn_j \sim \mathcal{CN}(\mathbf{0}, \sigma_j^2 \bmI)$ denote the noise with variance $\sigma_j^2$ at the DL user $j$. 
 Let $\bmR_j= \mathbb{E}[\bmy_j \bmy_j^H]$  denote the signal-plus noise covariance matrix at the DL user $j$. Let $\bmR_{\overline{j}} =  \mathbb{E}[\bmn_j \bmn_j^H]$ denote its noise covariance matrix. The rate at the DL user $j$ is given by
\vspace{-2mm}
\begin{equation}
    \mathcal{R}_j = \mbox{log} \big[\;\mbox{det}(\bmR_{\overline{j}}^{-1} \bmR_j) \big].
\end{equation}
\vspace{-1mm}
 
\subsection{Radar Model}

We consider the radar to estimate one angle of arrival (AoA), denoted as $\theta_k$. Let $\omega_0$ denote the angle between the FD BS and the RIS, which is perfectly known at the BS. Let $\bmy_b$ denote its total received signal, and let $\bmA$ denote a matrix, containing contributions from all the paths, defined as

\begin{equation}
\begin{aligned}
     \bmA =  &   \psi_k \bma_{r}(\theta_k) \bma_t(\theta_k)^T  + \xi_{1,k} \bma_{r}(\omega_0)  \bma_t(\omega_0)^T \mathbf{\Phi}    \bma_{i}(\theta_k) {\bma_{i}(\theta_k)}^T \\&\mathbf{\Phi} 
    \bma_{i}(\omega_0) \bma_t(\omega_0)^T
      + \xi_{2,k}\; \bma_{r}(\theta_k) \;{\bma_{i}(\theta_k)}^T \mathbf{\Phi}    \bma_{i}(\omega_0) \bma_t(\omega_0)^T  \\& + \psi_{b,i}  \bma_{r}(\omega_0)  \bma_t(\omega_0)^T  \mathbf{\Phi}  \xi_{3,k} \;\bma_{i}(\theta_k)\;  {\bma_{t}(\theta_k)}^T,
\end{aligned}
\end{equation}
where $\psi_k$ denotes the reflection coefficient for the LoS path between the BS and target $k$, $\bmn_b \sim \mathcal{CN}(\mathbf{0}, \sigma_r^2 \bmI)$ denotes the noise at the radar with variance $\sigma_r^2$, and $\bma_r$ and $\bma_t$ denote the transmit and receive antenna steering vectors for JCAS node, respectively. The scalars $\xi_{1,k},\xi_{2,k}$ and $\xi_{3,k}$ denote the reflection coefficients for the signal from RIS to target $k$ and back to RIS, from RIS to the radar via target $k$ and from the transmitter to the RIS via target $k$, respectively. Given $\bmA$, we can write the received signal $\bmy_b$ as 
\begin{equation} \label{radar_model}
\begin{aligned}
 \bmy_b = & \bmA \bmV_j \bms_j +  \mathbf{H}_{b,b}^{l} \bmV_j \bms_j  +  \bmH_{b,i} \mathbf{\Phi}\bmH_{i,b} \bmV_j \bms_j   + \bmn_b,
\end{aligned}
\end{equation}
which contains the effective SI. Assuming uniform linear arrays at the FD node, the antenna steering vector $\bma_r$ at the receiver, appearing in $\bmA$, can be modelled as \vspace{-1mm}
\begin{equation} \label{array_response_ULA}
        \bma_r(\theta_k) = \frac{1}{\sqrt{N_b}}[1, e^{j \frac{2 \pi}{\lambda} d sin(\theta_k)},...,e^{j  \frac{2 \pi}{\lambda} d (N_b-1) sin(\theta_k)}]^T, \vspace{-2mm}
\end{equation}
and similar modelling also holds for $\bma_{t}$. The scalars $d$ and $\lambda$ denote the distance and wavelength, respectively. Let $\phi_{k}$ and $\varphi_{k}$ denote the elevation and the azimuth angles between the RIS and the target $k$. The RIS response can be modelled as a uniform planer array (UPA) as 
\vspace{-1mm}
\begin{equation}
    \bma_{i}(\phi_{k},\varphi_{k}) = \frac{1}{\sqrt{R C}}[1, e^{j \frac{2 \pi}{\lambda} 	\varpi_1},...,e^{j \frac{2\pi}{\lambda}	\varpi_{RC-1}}]^T,
\end{equation}  \vspace{-1mm}
where $\varpi_i = d_{i_x} sin(\phi_{k}) cos(\varphi_{k}) + d_{i_z} sin(\varphi_{k})$ \cite{chen2022hybrid}, $0 < i < RC$, and $d_{i_x}$
and $d_{i_z}$ denote the distance of RIS's element $i$ from its first element on the $x$ and $z$ axis, respectively. As the position and the orientation of RIS are known at the FD BS, the azimuth and elevation angles can be expressed as a function of $\theta_k$ to be estimated as
\begin{subequations}
    \begin{equation}
        \phi_{k} =   arcos\big(\frac{l_k cos(\theta_k) - r_{1} cos(\omega_0)}{r_{2}}\big),
\end{equation}
\begin{equation}
        \varphi_{k} =  arcos(\frac{l_k cos(\theta_k) - r_1 cos(w_0)}{r_r cos(\theta_k)}\big),
    \end{equation}
\end{subequations}
where $l_k$ is the distance of the target $k$, $\omega_0$ is the angle between the FD BS and RIS, and the scalars 
$r_1,l_k,r_2,r_r$, assuming the RIS to be placed on the (x,z) plane,  denote the distance between the FD BS and RIS with the relative angle $\omega_0$, the distance between RIS and target $k$, the distance between RIS and target $k$ on the (x,y) plane and the distance on the (x,z) plane, respectively.

To achieve an accurate estimation, the CRBs for AoA $\theta_k$ should be below the threshold $\zeta_k$, which imposes the constraint 
\begin{equation}\label{crb_constraint}
    \frac{1}{\zeta_k}  \leq \frac{1}{\mbox{CRB}(\theta_k)}, 
\end{equation}
where $\mbox{CRB}(\theta_k)$ is derived in the Appendix. 



\subsection{Problem Formulation}

We embark on the task of jointly optimizing the performance of both communication and radar systems. While the concept of communication rate is well-defined for the DL user, there is no such concept that exists for the UL.
The sum rate maximization problem can be formulated as a function of MMSE 
\cite{christensen2008weighted}. Conversely, even though the notion of UL rate in the context of FD JCAS is not well-established, the challenge of SI cancellation can be framed as a minimization problem, specifically minimizing the squared Frobenius norm of the SI power, thereby improving the accuracy of radar detection.

For the DL user, we assume that it deploys the combiner $\bmF_j$ to estimate its data streams $\bms_j$ as
$ \hat{\bms}_j = \bmF_j \bmy_j$. Assuming that the combiner $\bmF_j$ is optimized based on the minimization of the MSE criteria, its closed-form solution is
\begin{equation} \label{comb}
\begin{aligned}
     \bmF_j = & \bmV_j^H (\bmH_{j,b} + \bmH_{j,i} \mathbf{\Phi} \bmH_{i,b})^H ((\bmH_{j,b} + \bmH_{j,i} \mathbf{\Phi} \bmH_{i,b}) \bmV_j \bmV_j^H \\&(\bmH_{j,b} + \bmH_{j,i} \mathbf{\Phi} \bmH_{i,b}) + \sigma_r^2 \bmI)^{-1}.
\end{aligned}
\end{equation}
Given $\bmF_j$ as \eqref{comb}, the MSE of the DL user $j$ becomes

\begin{equation}
    \bmE_j = (\bmI + \bmV_j^H  \bmH_j^H \bmR_{\overline{j}}^{-1}   \bmH_j \bmV_j)^{-1}, 
\end{equation}
 where $ \bmR_j = {\sigma_j^2} \bmI$ is the noise covariance matrix with variance $\sigma_j^2$. Let $\bmW_j$ denote the weight computed as \cite{christensen2008weighted}
\begin{equation} \label{weights}
    \bmW_j = \frac{w_j}{\mbox{ln}2} \bmE_j^{-1}.
\end{equation}

The joint minimization problem under the total sum-power, CRB and unit-modulus constraint for the RIS can be stated as 

\begin{subequations} \label{org_cst}
\begin{equation}    \underset{\substack{\bmV_j,\mathbf{\Phi}}} {\min} \quad  \mbox{Tr}(\bmE_{SI}) + \mbox{Tr}(\bmW_j \bmE_j)
    \end{equation}
\begin{equation} \label{c1_WSR}
\text{s.t.} \quad  \mbox{Tr}\big(\bmV_{j} \bmV_{j}^H \big) \leq  p_o,
\end{equation}
\begin{equation} \label{c2_WSR}
  \quad    |\mathbf{\phi}(i)| = 1, \forall i,
\end{equation}
\begin{equation} \label{c3_WSR}
    \eqref{crb_constraint}
\end{equation}
 \end{subequations}
where $\bmE_{SI}$ is a matrix defined as 
\begin{equation}
\begin{aligned}
      \bmE_{SI} = & \bmH_{b,b}^l \bmV_j \bmV_j^H {\bmH_{b,b}^l}^H + \bmH_{b,b}^l \bmV_j \bmV_j^H \bmH_{i,b}^H \mathbf{\Phi}^H \bmH_{b,i}^H  + \bmH_{b,i} \mathbf{\Phi} \\& \bmH_{i,b} \bmV_j \bmV_j^H {\bmH_{b,b}^l}^H + \bmH_{b,i} \mathbf{\Phi} \bmH_{i,b} \bmV_j \bmV_j^H  \bmH_{i,b}^H \mathbf{\Phi}^H \bmH_{b,i}^H
\end{aligned}
\end{equation}
obtained by writing the Frobenius norm squared as a function of the trace operator, whose diagonal elements capture the effective SI power and \eqref{c1_WSR} and \eqref{c2_WSR} denote the total sum-power constraint $p_0$ and the unit-modulus constraint on RIS.
 
\section{Novel Algorithm Design for FD JCAS} \label{algo}

In this section, we provide a novel algorithm to solve the optimization problem \eqref{org_cst} based on alternating optimization.
Let $\mathcal{L}$ denote the Lagrangian of \eqref{org_cst} and let $\lambda_0$ and $\mu_k$ denote the Lagrange multipliers for the total sum-power constraint at the FD BS and for the CRB constraint of target $k$, respectively.

\subsection{Digital Beamformer Optimization}
 
 To optimize the digital beamformer $\bmV_j$, which jointly optimizes the DL rate and handles the SI for JCAS, we take the derivative of $\mathcal{L}$ with respect to the conjugate of $\bmV_j$, which leads to the following optimal WMMSE digital beamformer

\begin{equation} \label{digital_BF}
\begin{aligned}
      \bmV_j = & \Big( (\bmH_{j,b} + \bmH_{j,i} \mathbf{\Phi} \bmH_{i,b})^H \bmF_j^H \bmW_j \bmF_j (\bmH_{j,b} + \bmH_{j,i} \mathbf{\Phi} \bmH_{i,b}) \\& +  (\mathbf{H}_{b,b}^{l} + \bmH_{b,i} \mathbf{\Phi}\bmH_{i,b})^H (\mathbf{H}_{b,b}^{l} + \bmH_{b,i} \mathbf{\Phi}\bmH_{i,b}) \\& + \mu_k 2 \bar{\bmA}_{\theta_k}^H \mathbf{\Sigma}^{-1} \bar{\bmA}_{\theta_k} + \lambda_0 \bmI\Big)^{-1} \hspace{-2mm} (\bmH_{j,b} + \bmH_{j,i} \mathbf{\Phi} \bmH_{i,b}) \bmF_j^H \bmW_j.
\end{aligned}
\end{equation}
where $\bar{\bmA}_{\theta_k}^H$ is defined in the Appendix.
To find the optimal values of $\lambda_0 $ and $\mu_k$, a linear search method can be adopted. In this study, we employ the Bisection method.

 \subsection{RIS Optimization}

We consider optimizing the RIS to jointly achieve quasi-ideal SIC on the UL side (which can be seen as a virtual UL rate optimization problem) and also to maximize the rate at the DL side. For such a purpose, we consider minimizing the Frobenius Norm squared of the effective SI. Let $\bmB,\bmC,\bmD$ be
\begin{subequations}
\begin{equation}
     \quad \bmB = \bmH_{b,i}^H \bmH_{b,i} +  \bmH_{j,i}^H \bmF_j^H \bmW_j \bmF_j \bmH_{j,i},
\end{equation}
\begin{equation}
    \bmC = \bmH_{i,b}  \bmH_{i,b}^H + \bmH_{i,b} \bmV_j \bmV_j^H \bmH_{i,b}^H,
\end{equation}
   \begin{equation}
    \bmD = \bmH_{i,b} {\bmH_{b,b}^l}^H \bmH_{b,i} + \bmH_{i,b} \bmV_j \bmV_j^H \bmH_{j,b}^H \bmF_j^H \bmW_j \bmF_j \bmH_{j,i},
\end{equation}
\end{subequations}

 Given the auxiliary matrices, problem \eqref{org_cst}
can be restated with respect to diagonal elements of RIS $\mathbf{\phi}$ as

 \begin{subequations}  \label{eq_finale}
\begin{equation}    \underset{\substack{\mathbf{\phi}}} {\text{min}} \quad  \mathbf{\phi}^H \mathbf{\Lambda} \mathbf{\phi} + \bmd^T \mathbf{\phi} + \mathbf{\phi}^H  \bmd^*
    \end{equation}  \vspace{-3mm}
   \begin{equation}
       \eqref{c2_WSR} 
   \end{equation}   
\end{subequations} 
where $\mathbf{\Lambda} = (\bmB \odot \bmC )$, and $\bmd$ is a vector made of the diagonal elements of the matrix $\bmD$. To render a feasible solution, we adopt the majorization-maximization optimization method \cite{pan2020multicell} by constructing an upper bound denoted as $g(\cdot)$ for the objective function \eqref{eq_finale}, denoted as $f(\cdot)$. For a problem of type \eqref{eq_finale}, it has been shown in \cite{pan2020multicell} that at iteration $n$ the following upper bound can be considered
 \begin{equation} \label{UB}
     g(\bm{\phi}|\bm{\phi}^{(n)}) = 2 \mbox{Re}\{{\bms}^H \bmq^{(n)}\} + c, 
 \end{equation}
 where $\lambda^{max}$ is the maximum eigenvalue  $\bmq^{(n)} = (\lambda^{max} \bmI - \mathbf{\Lambda}) \bm{\phi}^{(n)} - {\bmd}^*$. Given the upper bound and $\bmq^{(n)}$, our problem can be restated as a minimization of the upper bound as, 
\begin{subequations} \label{restated}
   \begin{equation}
    \underset{\substack{\bm{\phi}_r}}{\min} \quad  2 \mbox{Re}\{{\bmd}^H \bmq^{(n)}\},
    \end{equation}
\begin{equation}
\text{s.t.} \quad 
   |\bm{\phi}(i)| = 1,  \quad \forall i,
   \end{equation}
\end{subequations}
By solving problem \eqref{restated}, we get the following 
\begin{equation} \label{solution_phi}
    \bm{\phi}^{(n+1)} = e^{i \angle\bmq^{(n)}}.
\end{equation}
When the digital beamformer is computed under the CRB constraint at each iteration, the RIS optimization should be carried out by solving the aforementioned problem iteratively until convergence. The procedure for optimizing the phase response of RIS is given in Algorithm~$1$. The overall procedure to optimize and solve the joint optimization problem under the CRB is formally given in Algorithm $2$.

 \begin{algorithm}[t]  
\caption{Optimization of RIS}\label{alg_1}
\textbf{Initialize:}  iteration index $n=1$, accuracy $\epsilon$.\\
\textbf{Evaluate:} $f(\bm{\phi}(0))$.\\
\textbf{Repeat until convergence}
\begin{algorithmic}
\STATE \hspace{0.001cm} Calculate $\bmq^{(n)} = (\lambda_i^{max} \bmI - \mathbf{\Lambda}) \bm{\phi}^{(n)} - {\bmd}^*$
\STATE  \hspace{0.001cm} Update $\bm{\phi}^{(n+1)}$ as $\bm{\phi}^{(n+1)} = e^{i \angle\bmq_i^{(n)}}$.\\
\STATE  \hspace{0.001cm} \textbf{if} $|f(\bm{\phi}_i^{(n+1)}) - f(\bm{\phi}_i^{(n)})|/f(\bm{\phi}_i^{(n+1)}) \leq \epsilon$
\STATE  \hspace{0.4cm} Stop and return $\bm{\phi}^{(n+1)}$.
\STATE  \hspace{0.001cm} \textbf{else} n=n+1 and repeat.
\end{algorithmic} \label{Alg_1}
\end{algorithm}

\begin{algorithm}[t]  
\caption{JCAS Optimization with SIC}\label{alg_3}
\textbf{Initialize} iteration index $n$, accuracy $\epsilon$, digital beamformer and combiner.\\
\textbf{Repeat until convergence}
\begin{algorithmic}
\STATE \hspace{0.3cm} Update $\bmF_j$ with \eqref{comb}.\\
\STATE \hspace{0.3cm} Update $\bmW_j$ with \eqref{weights}.\\
\STATE \hspace{0.3cm} Update $\bmV_j$ with \eqref{digital_BF}.\\
\STATE \hspace{0.3cm} Search for the Lagrange multipliers.
\STATE \hspace{0.3cm} Update $\mathbf{\Phi}_i$ with Algorithm  \ref{Alg_1}.\\
\STATE  \textbf{if} convergence condition is satisfied\\
\STATE  \hspace{0.4cm} Stop and return the optimized variables.
\STATE  \textbf{else} repeat.
\end{algorithmic}
\end{algorithm}  
The convergence of the proposed scheme is straightforward by combining the reasoning of the well-established WMMSE \cite{christensen2008weighted} and the majorization-maximization technique \cite{pan2020multicell}. However, due to space limitations, we omit the extended proof.

\section{Simulation Results} \label{risultati}

In this section, we present simulation results to validate the advantage of the proposed SI-aware FD JCAS transceiver design. 

We consider the FD BS and the DL user to be equipped with uniform linear arrays (ULA) at $80$~m from the FD BS, and the FD BS to be placed in the center of the three-dimensional coordinate system with ULA aligned with the z-axis. The RIS is assumed to be placed on the (x,y) plane, with a relative angle of $30^{\circ} $ with respect to the FD BS, with its first element being $5~$m far from the first transmit antenna of the FD BS. We assume that the FD BS is assumed to be equipped $M_b = 15$ transmit antennas and $N_b =10$ receive antennas, and the DL user is assumed to be equipped with $N_j = 5$ receiving antennas. The RIS is assumed to be of size $10 \times 10$. The channels between the FD BS and the DL user $j$, denoted with $\bmH_{j,b}$ and $\bmH_{b,i}$, are modelled with the line of sight (LoS) channel model. The number of data streams to be transmitted to the DL user is set to be $d_j =2$. The digital beamformer $\bmV_j$ is initialized as the dominant eigenvectors of the effective channel covariance matrices, and the response of the NF-IRSs is initialized with random phases.
We define the signal-to-noise-ratio (SNR) of our system as
$ SNR = p_o/\sigma_j^2$, where $p_o$ is the total transmit power and $\sigma_j^2$ is the noise variance at the DL user. For the CRB constraint, we set $\zeta_k = 0.01$ and the AoA to be estimated to be randomly distributed on a circle of 50$m$ with the angle range limited to $[-\pi/2,-\pi/2]$.

For comparison, we define the following benchmark schemes: 1) \emph{RIS-Communications Only} - A scheme in which the beamformers and RIS are designed to maximize the performance of the communications and there is no sensing and SI (half-duplex (HD) mode), 2) \emph{No RIS-Communications Only} - A scheme similar to scheme 1) but without RIS, and 3) \emph{No RIS- With Sensing}- A scheme in which FD BS performs JCAS but without the aid of RIS. We label our scheme as \emph{RIS-With Sensing}.

\begin{figure} 
    \centering
\includegraphics[width=0.8\columnwidth,height=5cm]{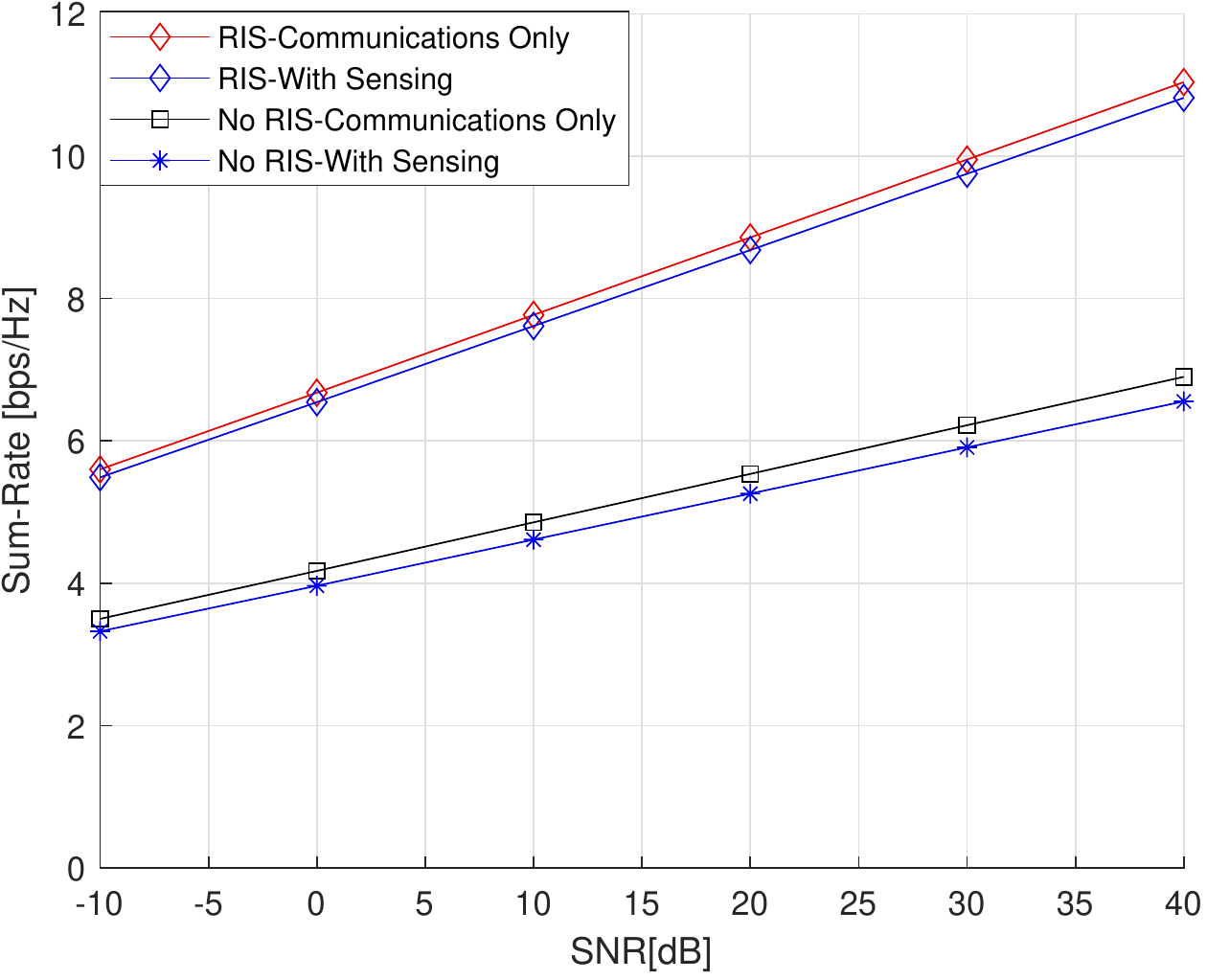}
    \caption{Communications performance for FD JCAS with SIC.}
    \label{rate_performance} \vspace{-2mm}
\end{figure}

\begin{figure}
    \centering
\includegraphics[width=0.8\columnwidth,height=5cm]{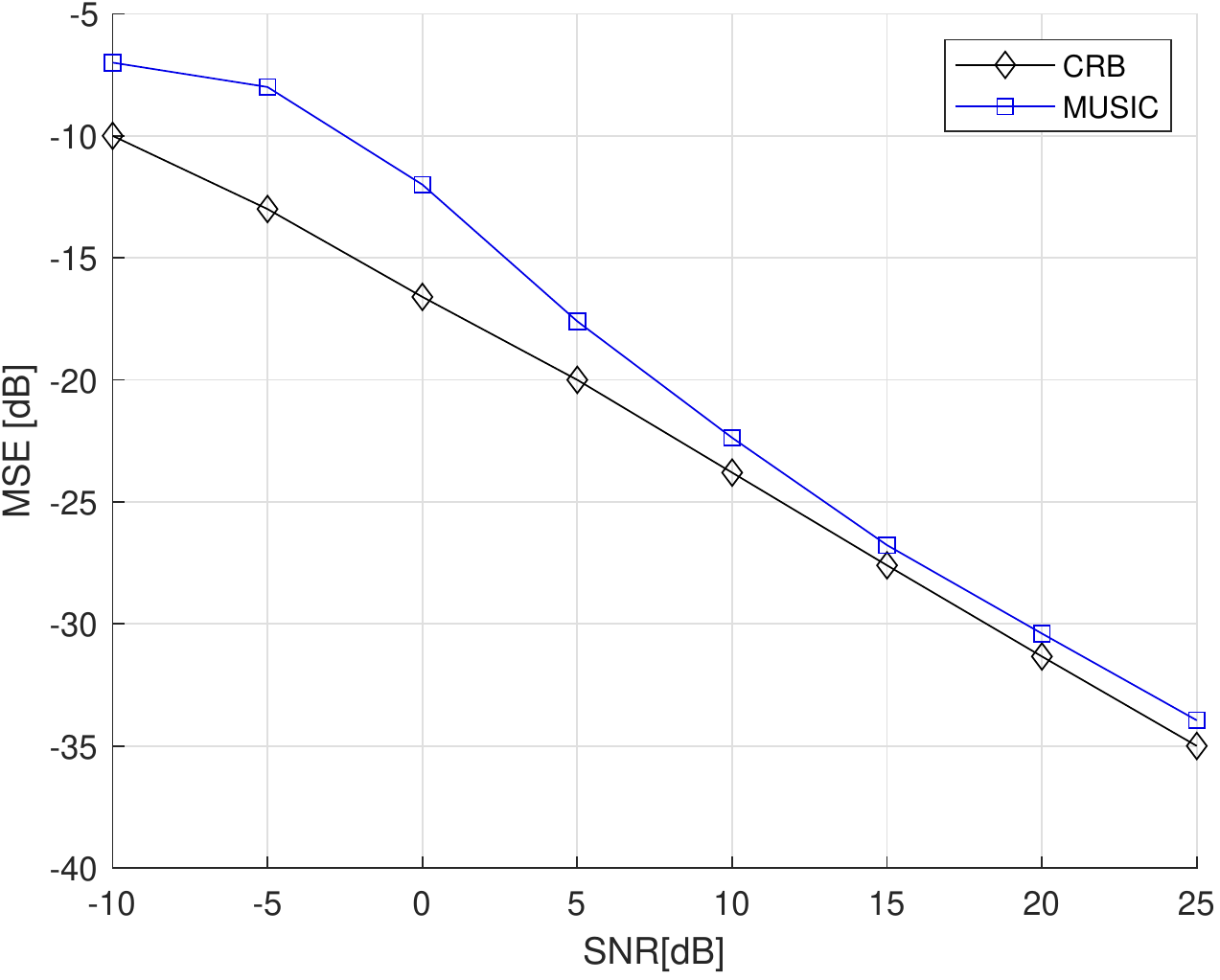}
    \caption{MSE for the AoA estimation as a function of SNR.}
    \label{fig2} \vspace{-2mm}
\end{figure}

Fig. \ref{rate_performance} shows the performance of the communications in terms of the user rate as a function of the proposed novel FD JCAS transceiver design, in comparison to the benchmark schemes. We can see that when the BS act as an HD BS and there is no SI and sensing, the beamformer $\bmV_j$ and the RIS is designed to only maximize the DL performance, leading to a higher rate. However, in the FD JCAS case, the beamformers and RIS are designed to jointly handle SI, enhance the sensing performance and improve the communications.  

In Fig. \eqref{fig2}, we compare the performance of the estimation of AoA $\theta_k$ when using a MUSIC-based estimator. Our approach demonstrates effective management of SI, and as the SNR increases, the estimation performance approaches the CRB. While there remains a slight difference between the estimation performance and the CRB, this gap can be further reduced by incorporating the CRB constraint into the optimization of the Reconfigurable Intelligent Surface (RIS), which is a direction for future research.

\section{Conclusions} \label{conclusioni}
This work introduces a new method to enable FD JCAS by considering the impact of SI. The authors derive the exact CRB for RIS-assisted JCAS and propose a joint optimization framework based on alternating optimization that satisfies the CRB constraint. Simulation results demonstrate that the proposed beamforming method, which accounts for SI, leads to a substantial performance improvement and effectively manages SI. Furthermore, the reduction in data transmission rate compared to a communications-only approach is negligible, which paves the path toward FD JCAS with accurate and energy-efficient SI management with RIS. 

\appendix
In this section, we derive the CRB for AoA $\theta_k$, which can be written as \cite{cheng2021transmit}

\begin{equation} \label{derived_CRB}
    \mbox{CRB}(\theta_k) = \frac{1}{ 2} \Big( \mbox{Tr}\big(\bmV_j^H \bar{\bmA}_{\theta_k}^H \mathbf{\Sigma}^{-1} \bar{\bmA}_{\theta_k} \bmV_j\big) \Big)^{-1}.
\end{equation}
where $\bar{\bmA}_{\theta_k} = \partial\bmA/ \partial \theta_k$.
We first define the derivatives of the antenna responses as

\begin{subequations}
   \begin{equation}
      \partial \bma_{r} = \frac{1}{\sqrt{N_b}}[0, ....,j \frac{2 \pi}{\lambda} d (N_b-1)cos(\theta_k) e^{j  \frac{2 \pi}{\lambda} d (N_b-1) sin(\theta_k)}]^T,
    \end{equation}
    \begin{equation}
      \partial \bma_{t} = \frac{1}{\sqrt{M_b}}[0,....,j \frac{2 \pi}{\lambda} d (M_b - 1 )cos(\theta_k) e^{j  \frac{2 \pi}{\lambda} d (M_b-1) sin(\theta_k)}]^T,
    \end{equation}
\end{subequations}
Let $\partial \varpi_i$ denote the derivative of $\varpi_i$, obtained by expressing $\phi_{k}$ and $\varphi_{k}$
as a function of $\theta_k$.
Then the derivative of $\bma_{r}^{RIS}$ with respect to $\theta_k$ can be written as
\begin{equation}
       \partial \bma_{t}^{RIS} = \frac{1}{\sqrt{RC}}[0,......, j \frac{2 \pi}{\lambda} \partial \varpi_i e^{j \frac{2\pi}{\lambda}	\varpi_{RC-1}}]
   \end{equation} 
By considering the complete deployment effect of RIS, including both the LoS and non-LoS links, $\bar{\bmA}_{\theta_k}$ can be written as
\begin{equation}
\begin{aligned}
     \bar{\bmA}_{\theta_k}  = & \psi_k \partial \bma_{r} (\theta_k) \bma_t(\theta_k)^T  + \psi_k \bma_{r}(\theta_k) \partial \bma_t(\theta_k)^T  \\& +  \xi_{1,k}  \bma_{r}(\omega_0)   \bma_t(\omega_0)^T  \mathbf{\Phi}    \partial \bma_{r}^{RIS}(\theta_k) {\bma_{i}(\theta_k)}^T \mathbf{\Phi} 
     \bma_{r}^{RIS}(\omega_0) \\&\bma_t(\omega_0)^T    +   \xi_{1,k}   \bma_{r}(\omega_0)  \bma_t(\omega_0)^T \mathbf{\Phi}  \bma_{i} (\theta_k)\partial {\bma_{i}(\theta_k)}^T \mathbf{\Phi} 
   \bma_{i}(\omega_0)  \\&  \bma_t(\omega_0)^T   + \xi_{2,k}\; \partial\bma_{r}(\theta_k) {\bma_{t}^{RIS}(\theta_k)}^T \mathbf{\Phi}   \bma_{i}(\omega_0) \bma_t(\omega_0)^T  \\&+ \xi_{2,k}\; \bma_{r}(\theta_k) \partial{\bma_{i}(\theta_k)}^T  \mathbf{\Phi}  \bma_{i}(\omega_0) \bma_t(\omega_0)^T \\& +  \xi_{3,k}  \bma_{r}(\omega_0)  \bma_t(\omega_0)^T  \mathbf{\Phi}    \partial \bma_{i}(\theta_k)\;  {\bma_{t}(\theta_k)}^T \\& + \xi_{3,k}  \bma_{r}(\omega_0)  \bma_t(\omega_0)^T  \mathbf{\Phi}   \bma_{i}(\theta_k)  \partial{\bma_{t}(\theta_k)}^T,
 \end{aligned}
 \end{equation}

 \section*{Acknowledgement}
  This work has been supported by the SNS JU TERRAMETA project under EU’s Horizon Europe research and innovation programme under Grant Agreement number 101097101.
\bibliographystyle{IEEEtran}
\bibliography{main}

\end{document}